\documentclass[twocolumn,aps,prb,superscriptaddress]{revtex4-2}

\usepackage{graphicx}
\usepackage{amssymb}

\begin{document}

\preprint{preprint}

\title{Hyperfine electro-nuclear coupling at the quantum criticality of YbCu$_4$Zn}



\author{S. Gabani}
\affiliation{
Institute of Experimental Physics, SAS, Watsonova 47 Košice, Slovakia}

\author{I. Čurlik}
\affiliation{Faculty of Sciences, University of Prešov, 17. novembra 1, SK - 080 78 Prešov, Slovakia}

\author{F. Akbar}
\affiliation{Department of Chemistry, University of Genova, Via Dodecaneso 31, Genova, Italy}

\author{M. Giovannini}
\affiliation{Department of Chemistry, University of Genova, Via Dodecaneso 31, Genova, Italy}

\author{J.G. Sereni}
\affiliation{Low Temperature Division, CAB-CNEA, CONICET, IB-UNCuyo, 8400 Bariloche, Argentina}

\email[]{jsereni@yahoo.com}

\date{\today}

\begin{abstract}

An increasing number of Yb-based compounds fulfill the conditions for the investigation of hyperfine electro-nuclear coupling effects related to $^{171}$Yb and $^{173}$Yb isotopes. Among them, the lack of magnetic order down to the milikelvin range in compounds with robust 
localized electronic moments and their nuclear magnetism. Although reminiscences of short range magnetic interactions may be observed below 1\,K, such perturbation can be dodged investigating compounds located close to a quantum critical point (QCP), where quantum fluctuations prevent the development 
of magnetic correlations to develop. 
Within the family of cubic YbCu$_4$M compounds (M = Ni, Au and Zn), we have investigated YbCu$_4$Zn that shows a logarithmic temperature dependence: $C_P/T 
\propto \ln(T/T_Q)$ in its electronic 
specific heat, as predicted for a QCP. Simultaneously, no signs of RKKY interactions are detected down to 0.03\,K. Due to the low Kondo temperature of its doublet 
ground state, the localized $4f$ electrons weakly couple with conduction electrons, allowing the coupling between nuclear and $4f$ electron moments to become relevant. 
However, the reminder Kondo interaction acts on the electronic hyperfine field producing a small deviation from the standard nuclear $C_N \propto 1/T^2$ 
dependence into a $n <2$ power law. The expected $n =2$ dependence is progressively recovered under applied magnetic field. 

\end{abstract}


\maketitle


\begin{figure}
\begin{center}
\includegraphics[width=18pc]{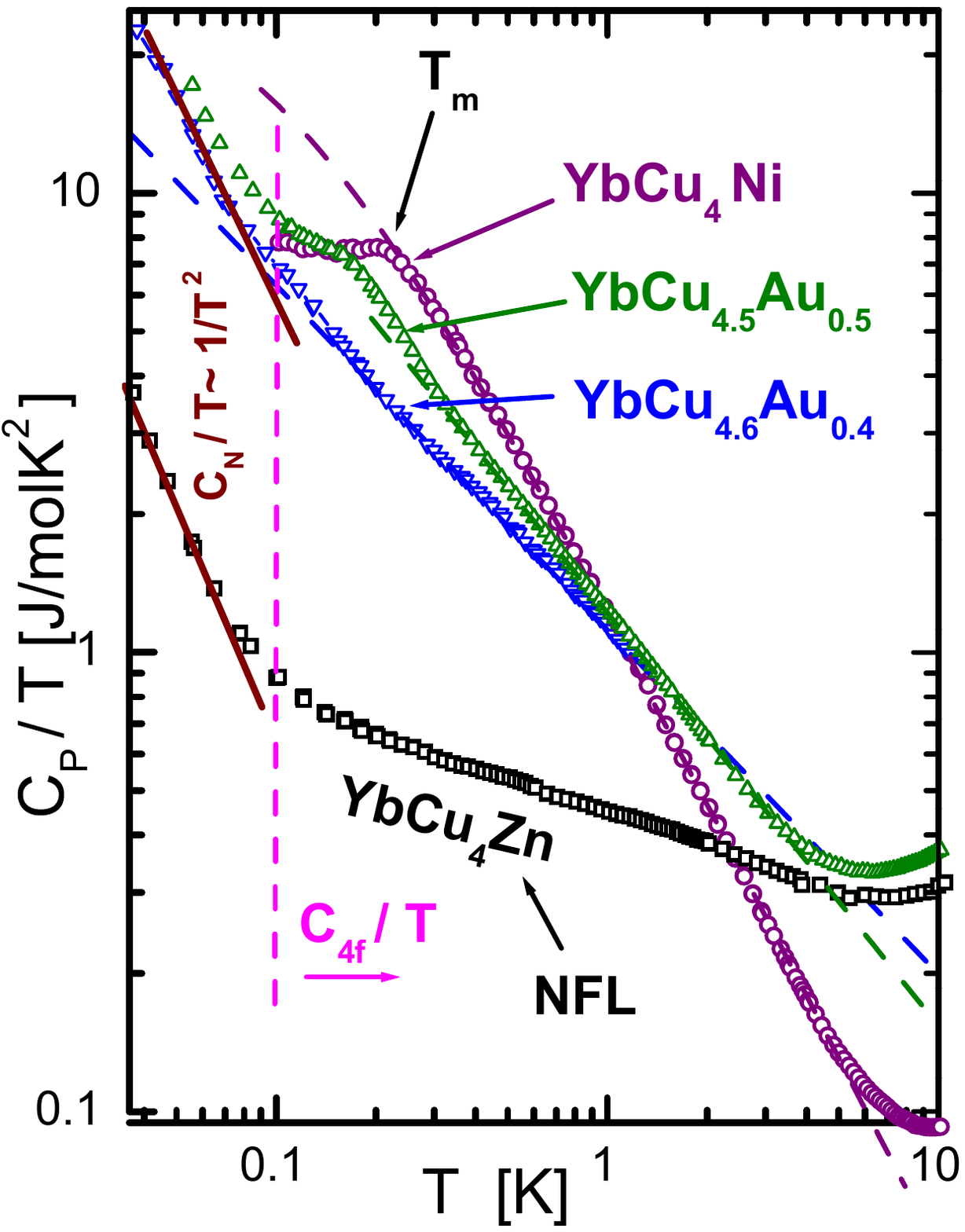}
\caption{(Color online) Comparison of the low temperature specific heat divided temperature of three YbCu$_4$M isoelectronic compounds lying in the vicinity of a 
QPC, in a 
double logarithmic representation. Dashed curves indicate respective power law fits. Note the coincident arising nuclear contribution $C_N/T$ below $T\approx 0.1$\,K, 
but with a $1/T^2$ dependence (brown lines). \label{F1}}
\end{center}
\end{figure}

\section{Introduction}

The basic requirements for a system to exhibit electronuclear hyperfine coupling are: to present a robust and localized (e.g. $4f$) magnetic moment without magnetic 
order, and to have isotopes with non zero nuclear momentum ($\vec I \neq 0$). The former condition can be achieved in systems tuned close to
a quantum critical point (QCP) \cite{QCP} where the magnetic order phase boundary extrapolates to zero or in presence of magnetic frustration \cite{SpinLiq}
The latter condition is 
fulfilled e.g. by Yb$^{3+}$ atoms whose natural concentration of $\vec I \neq 0$ isotopes is $14\%$ of $^{171}$Yb and  $16\%$ of $^{173}$Yb \cite{Frossati}. 

The family of cubic compounds YbCu$_4M$ ($M$ = Ni \cite{YbCu4Ni}, Au \cite{YbCu5-xAux}) and  Zn \cite{Serrao} and YbNi$_4N$ ($N$ = Mg 
\cite{YbNi4MgLatPar} and Cd \cite{Cd}) possess these characteristics. The 
topological conditions for magnetic frustration are provided by the fcc structure of MgCu$_4$Sn type (ternary derivative of the AuBe$_5$-type structure) with respective lattice constants a = 6.943\,\AA\,  
\cite{YbCu4Ni}, 7.046\,\AA\,\cite{YbCu4AuBau,YbCu5-xAux}, 7.046\,\AA\,\cite{Serrao}, 7.032\,\AA\,\cite{YbNi4MgLatPar} and 6.975\,\AA\, \cite{Cd} and full 
occupation of the atoms in each site. This fcc lattice can be viewed as 
a network of edge-sharing tetrahedra with Yb magnetic ions located at the vertices, being a 3D analog of a triangular lattice \cite{XtlChem, ActPol}. 

\subsubsection{Proximity to a QCP}

Although there is a larger number of YbCu$_4X$ compounds ($X$ = $M$ plus Cd, Ag, Tl \cite{Serrao} and In, Ga, Pd \cite{Ling}) with this cubic structure, only those belonging to the 
$M$ group were studied down to very low temperature. In  Fig.~\ref{F1} we compare the low temperature specific heat $C_P/T$ ratio of these compounds. 
Two different types of $C_P/T(T)$ dependencies can be distinguished in the figure.
One for $M$= Ni and Au (note that Ni results equal those of Au$_{0.6}$, therefore not included for clarity),  and the other for YbCu$_4$Zn.

In the case of YbCu$_{5-x}$Au$_x$ alloys, the decreasing 
transition temperature from $T_m(x) =0.8$\,K at $x=1$ towards a short range interactions crossover at $T_m(x=0.4) = 0.1$\,K \cite{YbCu5-xAux}, extrapolates to $T_m \to 
0$ for $x=0.32$ \cite{Banda}. Notably, the $T_m(x)$ anomaly seems to vanish before to rech $T_m =0$ like many other compounds approaching a QCP from the 
magnetic side \cite{PhilMg}. Coincidentally, these compounds show a 
power law $C_P \propto 1/T^p$ dependence above $T_m$ which is characteristic of a spin-liquid behavior \cite{SpinLiq}. Similar behavior is observed in YbCu$_4$Ni ($T_m = 0.2$\,K) \cite{YbCu4Ni}. In the case of isoelectronic Cu and Au atoms in the YbCu$_{5-x}$Au$_x$ alloys

Within the $x\leq 1$ range of concentration in the YbCu$_{5-x}$Au$_x$ isoelectronic Cu/Au alloys, Cu substitutes the larger Au atoms in the $4c$ Wyckoff 
position increasing the relative available volume \cite{availab} of Yb atoms. In fact, the lattice parameter decreases less than the extrapolation following the Vegards law from higher Au  
concentrations \cite{Arxiv}. Then, is the decrease of Au concentration that leads the decrease of the $T_m$ anomaly.

On the contrary, YbCu$_4$Zn does not show any magnetic anomaly and therefore it is placed on the non-magnetic side of the QCP. This positioning is not arbitrary 
because it is supported by a non-fermi-liquid (NFL) $C_P \propto log(T/T_Q)$ dependence, see the detail of these results in Fig.~\ref{F3}. 

\begin{figure}
\begin{center}
\includegraphics[width=22pc]{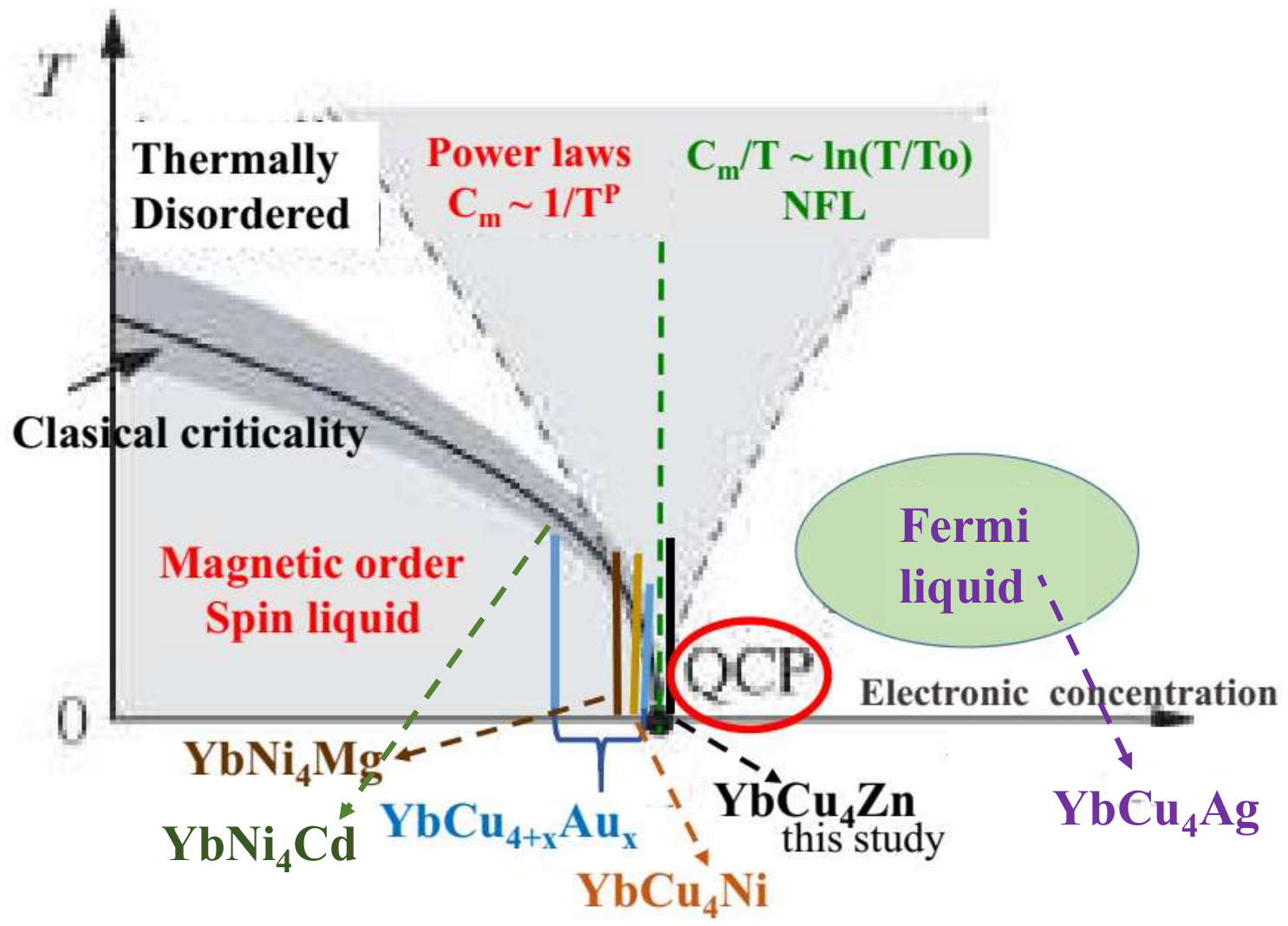}
\caption{(Color online) Schematic Donich-like magnetic diagram in the vicinity of a QCP,  comparing the relative positions of the isotypic Yb compounds as a function of their 
respective Fermi energies, see the text. \label{F2}}
\end{center}
\end{figure}

Taking profit of the vicinity of this isoelectronic family of compounds to a QCP, one can map their relative positioning according to the value of the respective magnetic 
anomaly at $T= T_m$ in a schematic Doniach phase diagram shown in Fig.~\ref{F2}. YbCu$_4$Zn is included according to its characteristic temperature 
$T_Q$ to be discussed in th following Section. 
The compound YbCu$_4$Ag is also included as a book example of Fermi liquid behavior, with the largest cell volume, lowest magnetization ratio (at $T=4$\,K) and 
electrical scttering \cite{Serrao} of these isotypic compounds.

Two Ni rich isostructural compounds are accounted for in Fig.~\ref{F2}, 
YbNi$_4$Mg ($T_m = 0.3$\,K) \cite{YbNi4Mg} and YbNi$_4$Cd ($T_m = 0.97$\,K) \cite{Cd}), as further examples for mapping this quantum critical region. 
Notably all these compounds can be ordered following the 'Electronic Concentration' as the unique parameter coupled to the mentioned available volume. It is known that Yb atoms have access to 
two electronic configurations: Yb$^{3+}$ and Yb$^{2+}$, being the former magnetic and the latter non-magnetic. Therefore, by increasing the electronic concentration of the ligand atoms 
and/or increasing the available volume of Yb cell the atom is progressively driven from its smaller Yb$^{3+}$ configuration  to the larger Yb$^{2+}$. For this family of compounds tuned close to a 

QCP it allows to investigate the evolution of the physical properties in the crossover of that critical point.

\subsubsection{Hyperfine electro-nuclear coupling}

Three different types of particles are involved in the hyperfine electro-nuclear coupling mechanisms in rare earth based intermetallic compounds: the 
conduction-electrons (with spin $\vec S$=1/2), the localized $4f$-electrons (mostly with $\vec J$=1/2 ground state after electronic crystal field plitting) and the nuclear moment 
(with orbital number $\vec 
I\neq 0$). 

The usual interactions between these particles are: i) the $4f$ magnetic RKKY interaction:     
{\it H}$_{R}$ = J$_{R}${\it (\^{J}*\^{J})}, ii) Kondo interaction between conduction and localized $4f$ electrons: {\it H$_K$} = J$_K${\it (\^{S}*\^{J})}, 
iii) the hyperfine interaction (hf)  
between $4f$ electrons and nuclear moments: {\it H$_{hf}$} = A$_{hf}${\it (\^{J}*\^{I})},  and iv) nuclear-dipolar interaction: {\it H$_{dip}$} = A$_{dip}${\it 
(\^{I}*\^{I})}.  The former, {\it H}$_{R}$, induces long 
range magnetic order that inhibits the $`hf'$ interaction, and the last is quite weak compared   with the other interactions. Quadrupolar interaction is not considered 
here because of the lack of electric gradient in the cubic 
crystallographic symmetry of the investigated compounds \cite{Mignot}. 

The effective magnetic field induced by the localized $4f$ electrons on the nuclear magnetic moments ($\mu_I=g_J\mu_n \vec I$) may reach hundreds of Tesla, that splits the $\vec I$ 
degenerated levels producing a low temperature Schottky anomaly in the nuclear specific heat $C_N(T)$. 

Therefore, this work is mainly devoted to the study of the competition between {\it H$_K$} and {\it H$_{hf}$} scales of 
energy because the former interaction triggers a fast relaxation mechanism ($\tau = h/k_BT_K$) that prevent the slower nuclear-electron relaxation \cite{Banda}. 

\section{Experimental results}

\subsection{Specific heat}

Going back to Fig.~\ref{F1}, the measured $C_P/T(T)$ of the three isotypic Yb-compounds can be better analyzed after defining two regions: the ''paramagnetic'' (for $T> 0.1$\,K), and the 
''nuclear'' one (for $T< 0.1$\,K), where the $C_N/T(T)$ component grows above the paramagnetic $C_{4f}/T(T)$ contribution.
In the former, one can appreciate that the compounds belonging to the `magnetic' side of the QCP follow a power law dependence: $C_P/T(T)\propto 1/T^p$, with an exponent $p=1.4$ for YbCu$_4$Ni \cite{YbCu4Ni}, that progressively decreases to 
$p=0.9$ for YbCu$_{4.5}$Au$_{0.5}$ \cite{YbCu5-xAux} and to $p=0.75$ for YbCu$_{4.6}$Au$_{0.4}$ \cite{Banda}. This confirms that YbCu$_{5-x}$Au$_{x}$ approches to a QCP from the magnetic side as Au concentration decreases. 

Notably, the specific heat anomalies at $T=T_m$ of YbCu$_4$Ni and YbCu$_{5-x}$Au$_{x}$ look like different stages in the evolution of the same type 
of magnetic crossover. 
In fact, the curve corresponding to YbCu$_{4.4}$Au$_{0.6}$ is not included in Fig.~\ref{F1} for clarity because it coincides with that of YbCu$_4$Ni. 

\begin{figure}
\begin{center}
\includegraphics[width=20pc]{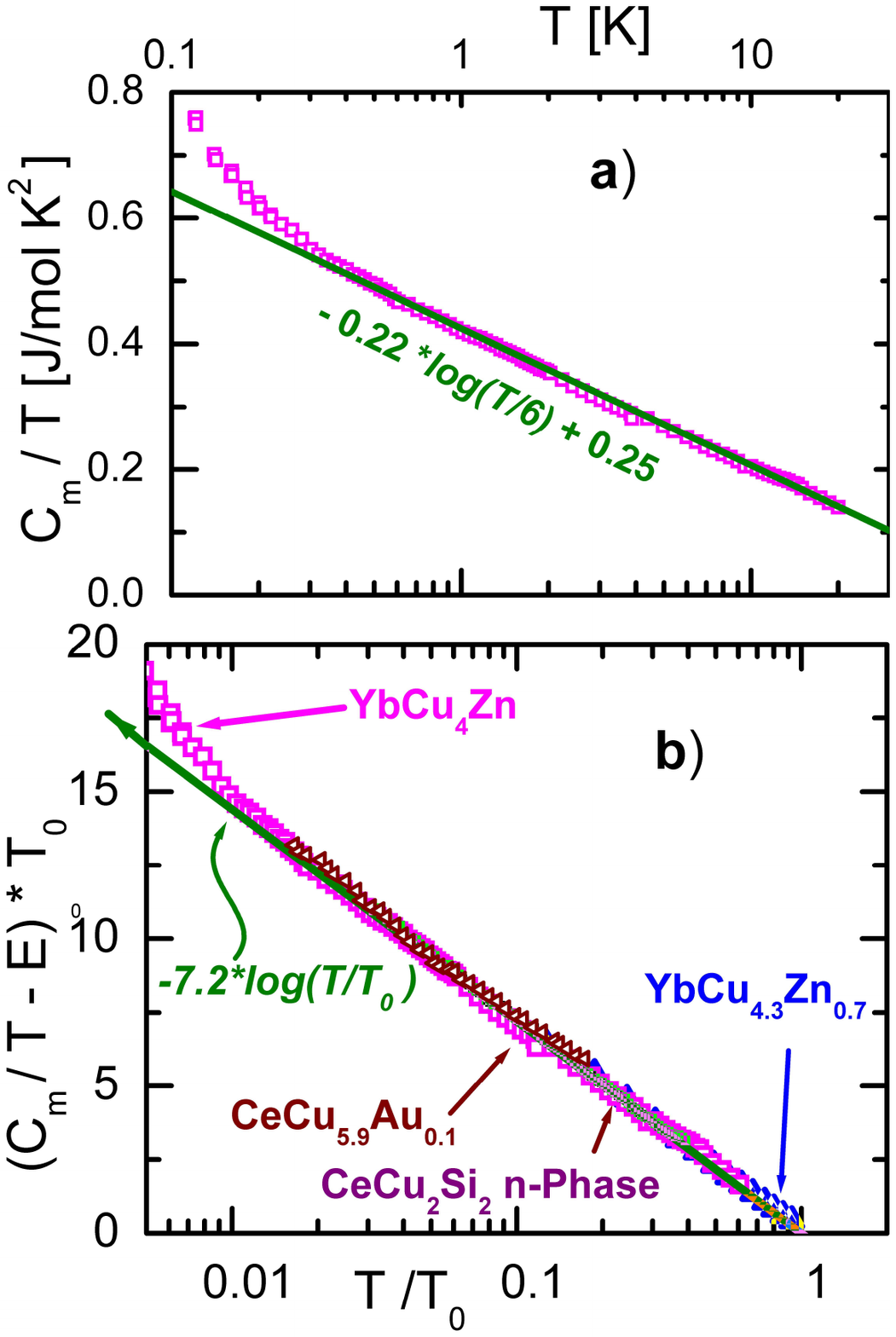}
\caption{(Color online) a) Logarithmic temperature dependence of  YbCu$_4$Zn. b) Comparison of different NFL compounds within the normalized procedure (see 
text). \label{F3}}
\end{center}
\end{figure}

As mentioned before, a completely different behavior of $C_P(T)/T$ is observed for YbCu$_4$Zn, that  corresponds to a logarithmic temperature dependence, as it is expected for non-fermi-liquid (NFL) systems \cite{Steward}. This compound is 
therefore placed into the 'non-magnetic' side of the QCP (see Fig.~\ref{F2}) because there are no traces of magnetic interactions, even under applied magnetic field of $B=9$\,T. The 
$\log(T/T_0)$ dependence is discussed in the next subsection devoted to the data presented  in Fig.~\ref{F3}a.

The comparison of the $C_P(T)/T$ dependencies of these compounds offers a unique opportunity to gain insight into the thermodynamic behavior of magnetic systems in the vicinity of a QCP 
because their isotypic and isostructural character. As it is presented in Fig.~\ref{F1}, the $T$ dependencies in the paramagnetic region indicates to which side of the QCP belongs the system. 
Following the decreasing evolution of 
the $p$ exponent of $C_P(T)/T\propto 1/T^p$, one notes that this function tends to $log(T/T_0)$ for very low $p$ values. We remark that the studied  YbCu$_{5-x}
$Au$_{x}$ alloys with $p=0.9$ and 0.75 are exceptional cases of $p\leq 1$ exponents which allow to speculate on the lack of a discontinuity undergoing a QCP because 
of the presence of quantum fluctuations. Therefore, the definition for a QCP as the limit of a 
second order transition extrapolation to zero \cite{QCP} does not occur in real systems because  their magnetic phase boundaries 
transform from well defined phase-transition into some type of `crossover' (at $T=T_m$) as the quantum fluctuations become dominant. Interestingly, in Fig.~\ref{F1} one can see that the degrees 
of freedom associated to the crossover vanish faster than $T_m\to 0$. This means that in this case quantum fluctuations blur the crossover preventing $T_m$ to reach $T=0$ as 
mentioned for other systems \cite{PhilMg}.

Concomitant with the vicinity to a QCP and the consequent lack of magnetic order both,   YbCu$_{4.6}$Au$_{0.4}$ alloy and stoichiometric YbCu$_4$Zn, clearly exhibit the arising nuclear contributions $C_N(T)/T$ below 
about 0.1\,K. Notably, in both cases $C_N(T)$ does not show the expected $C_N/T \approx \alpha  /T^3$ dependence \cite{Tari} for well defined nuclear levels. Instead, a  $
\alpha_n/T^{2+\epsilon}$ dependence is observed, as indicated on the low temperature side of Fig.~\ref{F1}. The origin of this deviation is discussed in Section IV.


\subsubsection{Analysis of the YbCu$_4$Zn results}

At low temperature, the logarithmic $C_m(T)$ behavior of YbCu$_4$Zn is demonstrated in Fig.~\ref{F3}a in a $C_m/T \propto log(T/T_Q)$ representation, where the characteristic 
temperature: $T_Q$, is a scale of energy for the quantum fluctuations. For this figure three ranges of specific heat measurements are matched: i) the very low temperature: $0.05 \geq 
T \geq 1$\,K, measured in a dilution refrigerator, 
ii) the low temperature $1 \geq T \geq 8$\,K measured in a PPMS devise (Quantum Design) and iii) the intermediate temperature $T\geq 8$\,K extracted form 
Ref. \cite{Serrao} after the $C_{ph}$ subtraction. The low $T_Q=6$\,K value, obtained from the fit in Fig.~\ref{F3}a, indicates the vicinity of this compound to the QCP. 

The NFL character of the $C_m(T)$ behavior of YbCu$_4$Zn can is verified using the universal scaling criterion for NFL systems: $C_m / t = -D log(t) + E T_0$ \cite{NFL}, as  
 shown in  Fig.~\ref{F3}b.
There, $t = T/T_0$ and $D = -7.2$\,J/molK. Since $D$ is fixed, $T_0$ is the only free parameter that provides an energy scale comparison between different systems. $E$ 
accounts for the conduction band 
contribution.  For YbCu$_4$Zn one obtains $T_0 = 33$\,K and $E= 90$\,mJ/mol\,K$^2$, that can be contrasted with CeCu$_{5.9}$Au$_{0.1}$ \cite{Cu59Au01}: $T_0 = 5.3$\,K and $E= 
53$\,mJ/Kmol, and the n-phase of CeCu$_2$Si$_2$ \cite{phaseN}: $T_0 = 14$\,K and $E=40$\,mJ/Kmol. Also for comparison, the Cu-rich alloy YbCu$_{4.3}$Zn$_{0.7}$ alloy 
\cite{Akbar} is included in the figure with $T_0 = 16$\,K and $E= 170$\,mJ/Kmol, suggesting to be closer to the QCP than the isotypic stoichiometric compound. In fact the 
standard $C_m(T)/T$ values measured above 2\,K are about 50\% larger than those of YbCu$_4$Zn. 

Although the $C_m/T \propto log(T/T_Q)$ dependence of YbCu$_4$Zn appeares as an exception within this family of compounds, it is not an exception because by 19\% of Sc substitution in Yb$_{1-x}$Sc$_x$Co$_2$Zn$_{20}$ this compound also shows a $log(T/T_Q)$ dependence approaching a QCP \cite{YbCo2Zn20}.

\subsection{Magnetic susceptibility} 

\begin{figure}
\begin{center}
\includegraphics[width=20pc]{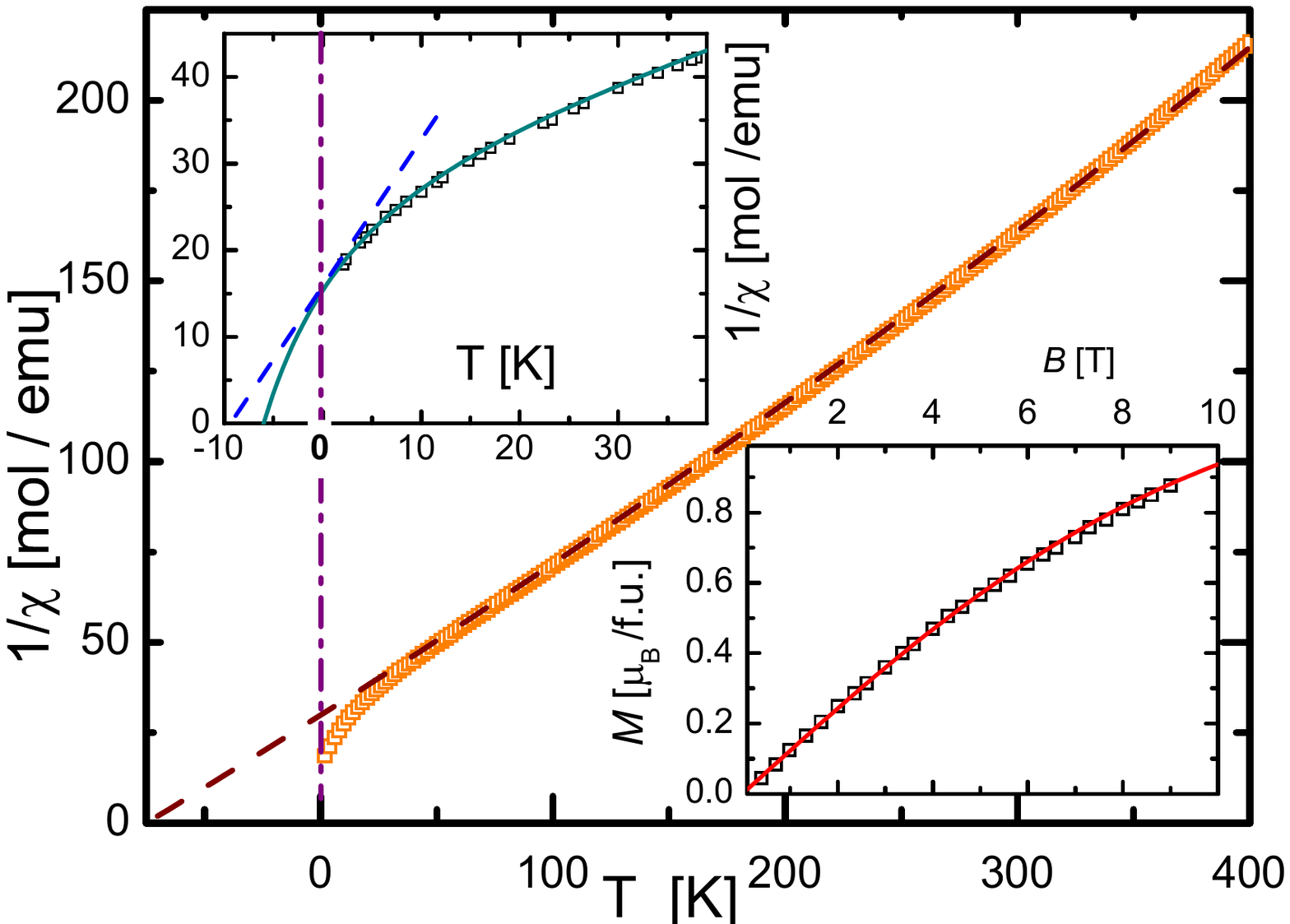}
\caption{(Color online) Inverse magnetic susceptibility of YbCu$_4$Zn measured at $B=1$\,T. Upper inset: detail at low temperature. Lower inset: Magnetization curve up to $B=9$
\,T at $T=2$\,K. The continuous curve is the fit with a Brillouin function $B_{1/2}(y)$, see the text. \label{F4}}
\end{center}
\end{figure}

The high temperature ($T\geq 50$\,K) inverse susceptibility of YbCu$_4$Zn is presented in Fig.~\ref{F4}. This measurement allows to determine an effective magnetic moment $
\mu_{eff} = 4.47 \mu_{B}$ (close to the theoretic value $4.54 \mu_{B}$ for $\vec{J}=7/2, \vec{S}=1/2,   \vec{L}=3$ with $g_J=1.14$)  and a paramagnetic temperature $\theta_P =-70$
\,K. From the relation $T_K = \theta_P/\sqrt 2$ \cite{Katanin} for the Kondo  temperature evluation, one extracts for high temperature $T_K^{HT} = 50$\,K that include the 
contribution of the excited CEF levels. 

High temperature specific heat measurements suggests the lowest CEF split is 
between two doublets with a gap of $\Delta_1 =35$\,K and therefore the $T\leq 30$\,K curvature of $1/\chi(T)$ in Fig.~\ref{F4} is expected to be related to the 
depopulation of the excited levels. In order to extract more information about the magnetic properties of the GS, we have fit the $1/\chi(T)$ dependence in the $T\leq 30$\,K range s 
presented in the upper inset of Fig.~\ref{F4}. 
The continuous curve fits the experimental data using the function:   
\begin{equation}
1/\chi(T)=(T+6)/ \Sigma_{i=0}^{i=2} \, a_i e^{-\Delta_i/(T+\theta_i)}
\end{equation}
where: $a_0=0.03$, $a_1=0.72$, $a_2=1.94$,  $\Delta_1 = 40\pm 5$\,K is close to the first CEF splitting and $\Delta_2 = 90\pm 15$\,K the energy of the quarted $\Gamma_8$, 
with $\theta_1 =25$\,K and $\theta_2 =41$
\,K as  
respective paramagnetic temperatures usually proportional to their Kondo scales. A $\theta_{GS}=-8.5\pm0.5$\,K is obtained for the GS as a linear extrapolation to $1/\chi =0$ from $T= 0$, i.e. following the tangent at $1/\chi (T=0)$ 
From this fit one can extract the relevant parameter $T_K=\theta_{GS}/\sqrt2= 6\pm0.3$\,K, which confirms the low Kondo scale of the GS and, from the slope of the $1/\chi (T=0)$ extrapolation, a $\mu_{eff}(GS)=2.13\mu_B$ is extracted. 

\subsection{Magnetization}

In the lower inset of Fig.~\ref{F4} we show the field dependent magnetization $M(B)$ measured at 2\,K. The experimental curve is properly fitted with a Brillouin function 
$B_{1/2}(y)$ (continuous curve), where  
$y = g_{eff} \mu_B J B/ k_B T$ and $g_{eff}  = 0.6$ for the present doublet GS. This function is normalized to a saturation value of $M_{sat} =1.23 \mu_B$.
Comparing this saturation value with the values for the two possible GS doublets: 
$M_{sat}(\Gamma_6) =1.33\mu_B$ and $M_{sat} (\Gamma_7) =1.72\,\mu_B$, it is evident that  $\Gamma_6$ is the most likely GS. Furthermore $\mu_{eff}(GS)/M_{sat}(2\,K) =2.13\mu_B/1.23 \mu_B=\sqrt 3$.
 

\section{Discussion}

Once established that YbCu$_4$Zn is a compound placed on the ''non-magnetic'' side of a QCP,   we can analyze the unexpected exponent of the power law 
increase of $C_P/T(T)$ below $T= 0.1$\,K, mentioned in Fig.~\ref{F1} within the context of the 
interaction between the localized $4f$ electrons and nuclear spins. 

From the comparison with the other isostructural compounds symmetrically placed on the ''magnetic'' side of the QCP one sees 
that: i) the so called Kondo interaction between local and band electrons ($T_K$) is present on  both sides of the QCP, ii) its energy scale is quite similar, concomitant with their similar electronic 
configurations and crystalline structures, and therefore: iii) there are no indications that $T_K\to 0$ at the QCP as it might be expected from theory, though the observed values are quite low and equivalent to those of quantum fluctuations. 

The question arises how these fluctuations: remnant of Kondo or quantum fluctuations, affect the interaction between local electrons and nuclear moments. For such a purpose we have 
studied the field dependence of the 
specific heat of YbCu$_4$Zn at vey low temperature and 
analyzed its evolution in the context of the hyperfine interaction.blundell

\subsection{Hyperfine coupling}

The dipole hyperfine coupling energy: $H_{hf} =  A_{hf}(\vec{\bf I}\cdot\vec{{\bf J}})$,  where $A_{hf}$ is the hyperfine constant, arises from the interaction between the nuclear  
$\vec{\bf I}$ and the electronic moments $\vec{\bf J}$. 

The element Yb (atomic number 70) possesses a number of stable isotopes \cite{Ybisotop}. Those with even mass, from 170 to 176 
and total abundance of $\approx 70\%$, have zero nuclear moment $\vec{\bf I}= 0$. The most relevant Yb isotopes with odd mass (171,173) and $\vec{\bf I}\neq 0$ nuclear moment have an abundance: 14 \% $^{171}$Yb 
with $\vec {\bf I} =1/2$ and  16 \% $^{173} $Yb with $\vec{\bf I} =5/2$ respectively.

Therefore, only the $\approx 30 \%$ isotopes with $\vec{\bf I} \neq 0$ are involved into the hyperfine coupling. As a consequence, the electronic magnetic properties acting on all Yb atoms in the paramagnetic region, is reduced to 1/3 of the atoms regarding the nuclear 
contribution observed at $T\leq0.1$\,K. This suggests that in the study of that contribution, 
the original chemical formula of YbCu$_4$ Zn can be reformulated for practical reason as: 
(0.7 $^{170}$Yb + 0.14 $^{171}$Yb + 0.16 $^{173}$Yb)Cu$_4$Zn below $\approx 0.1$\,K.  
Thus, only the 30\% of the sample: 0.14 $^{171}$Yb + 0.16 $^{173}$Yb)Cu$_4$Zn is involved in the hyperfine interaction.

The total angular moment originated in the combination of nuclear and electron moments ($H_{hf} =  A_{hf} ({\bf I*J}$), is given by: $\vec{\bf F} = \vec{\bf I} + \vec{\bf J}$ 
\cite{Blundell}, with the possible hyperfine couplings for each isotope: 
 i) $^{171}$Yb, with $\vec{\bf I}  =1/2$ and  $\vec{\bf J}  =1/2 => \vec{\bf F}$ = 0 and 1, and ii) $^{173}$Yb, with $\vec{\bf I}=5/2$ and  $\vec{\bf J}  =1/2 => \vec{\bf F}$ = 2 and 3 
The former case reminds the ortho-para nuclear spin order in Hydrogen molecules 
and magnetic dimers formation \cite{Blundell}.
We notice that in the isotropic cubic crystalline  symmetry of YbCu$_4$Zn, the nuclear Quadrupolar Gradients can be neglected \cite{Mignot}.  

\subsection{Nuclear contribution to specific heat}

\begin{figure}
\begin{center}
\includegraphics[width=20pc]{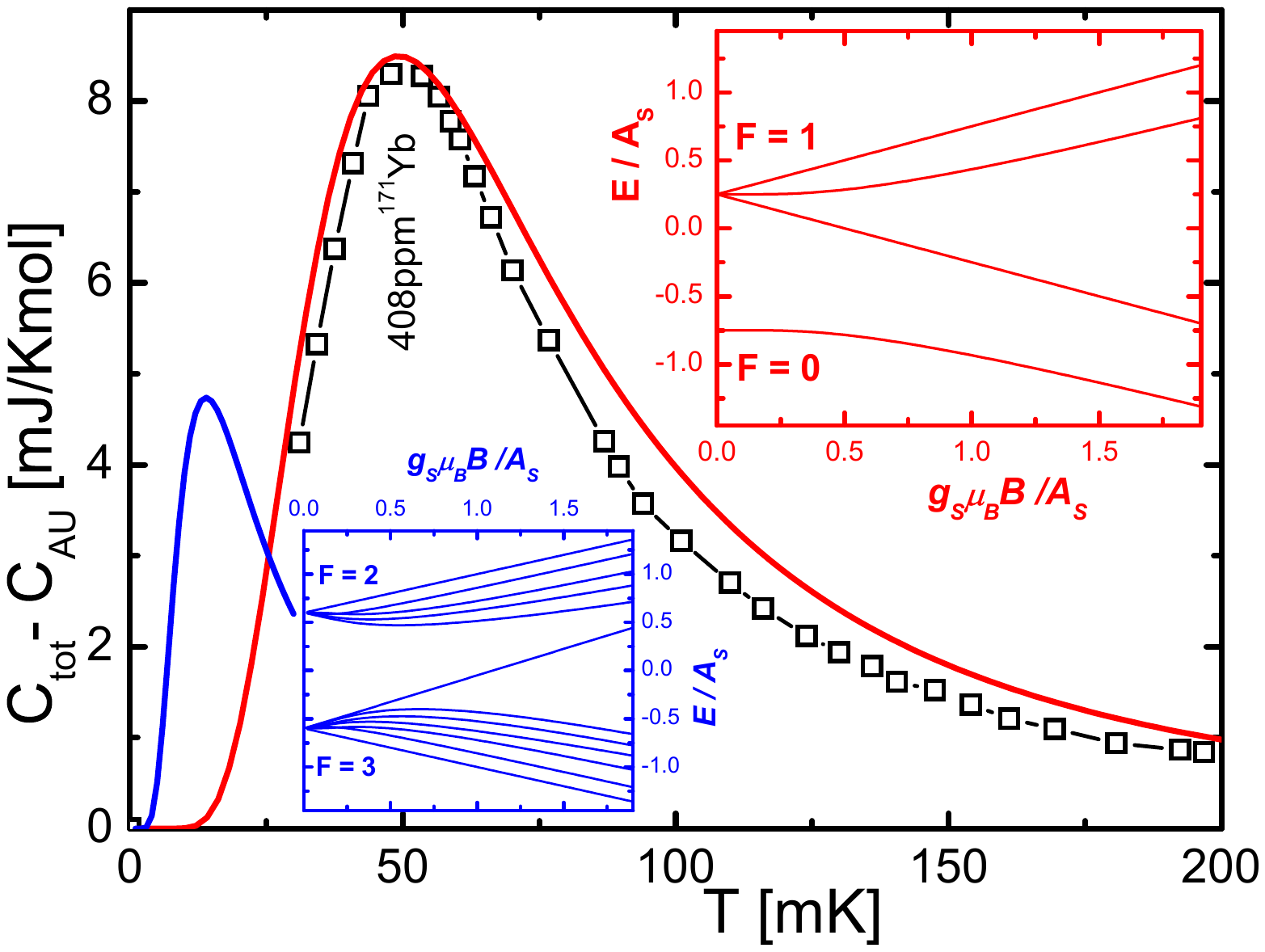}
\caption{(Color online) Nuclear contribution to specific heat of $^{171}Yb$ diluted in {\underline {Au}}{\it Yb} after \cite{Mignot} (open squares). Red curve: Schottky anomly fit following the energy levels  
scheme presented in the upper (red) inset, with ${\bf \vec F = \vec I \pm \vec J}$ = 1 and 0, from  $I=1/2$ and $J=1/2$, and $A_S > 0$. 
Blue curve: Schottky anomaly fit following the lower inset (blue) energy levels scheme, with ${\bf \vec F = \vec I \pm \vec J = 2, 3}$ from $I=5/2$ and $J=1/2$, and $A_{hf} < 0$. \label{F5}}
\end{center}
\end{figure}

In order to elucidate which one of the two contributions, from $^{171}$Yb or from $^{173}$Yb, is the  relevant in this case one can take as a reference the studies performed on diluted  $^{171}
$Yb into the cubic-fcc matrix Au \cite{Mignot}. 

In Fig.~\ref{F5}, the nuclear contribution to the specific heat of $^{171}$Yb: $C_N=C_{meas}- C_{Au}$, after \cite{Mignot} is prsented. Red curve: Schottky like fit with upper inset 
(red) level scheme: ${\bf \vec F = \vec I \pm \vec J}$ = 1 and 0, with $\vec I=1/2$ and $\vec J=1/2$, and $A_S > 0$. From this fit the `hf` factor $A_{1/2}(171) = 130$\,mK was extracted.  
Notice that in this configuration the GS is the singlet $\vec{\bf F_0} = 0$, while the excited state is the triplet $\vec{\bf F_1} = 1$

For comparison, the blue Schottky curve corresponds to the lower inset (blue) level scheme:  ${\bf \vec F = \vec I \pm \vec J }$= 2 and 3 with $\vec I=5/2$ and $\vec J=1/2$, and $A_{5/2} < 0$, with $A_{5/2}(173) =35$\,mK. There one can appreciate the difference in temperature and height of the maximum. It is evident that the envolved entropies are quite different in both energy 
schemes: with $\Delta S_1 = Rln(4)$ for $\vec{\bf F_0}\to \vec{\bf F_1}$, and $\Delta S_2 = 
Rln(12/7)$ for $\vec{\bf F_3} \to \vec{\bf F_2}$, even though the $A_{5/2}$ contribution is not negligible. However, for practical reasons, we will refer all the measured $C_N(T)$ contribution to 
the $A_{1/2}$ coupling scheme because it is the dominant one within our experimental range of temperature: $T\geq 50$\,mK.

\subsubsection{The case of YbCu$_4$Zn}

\begin{figure}
\begin{center}
\includegraphics[width=20pc]{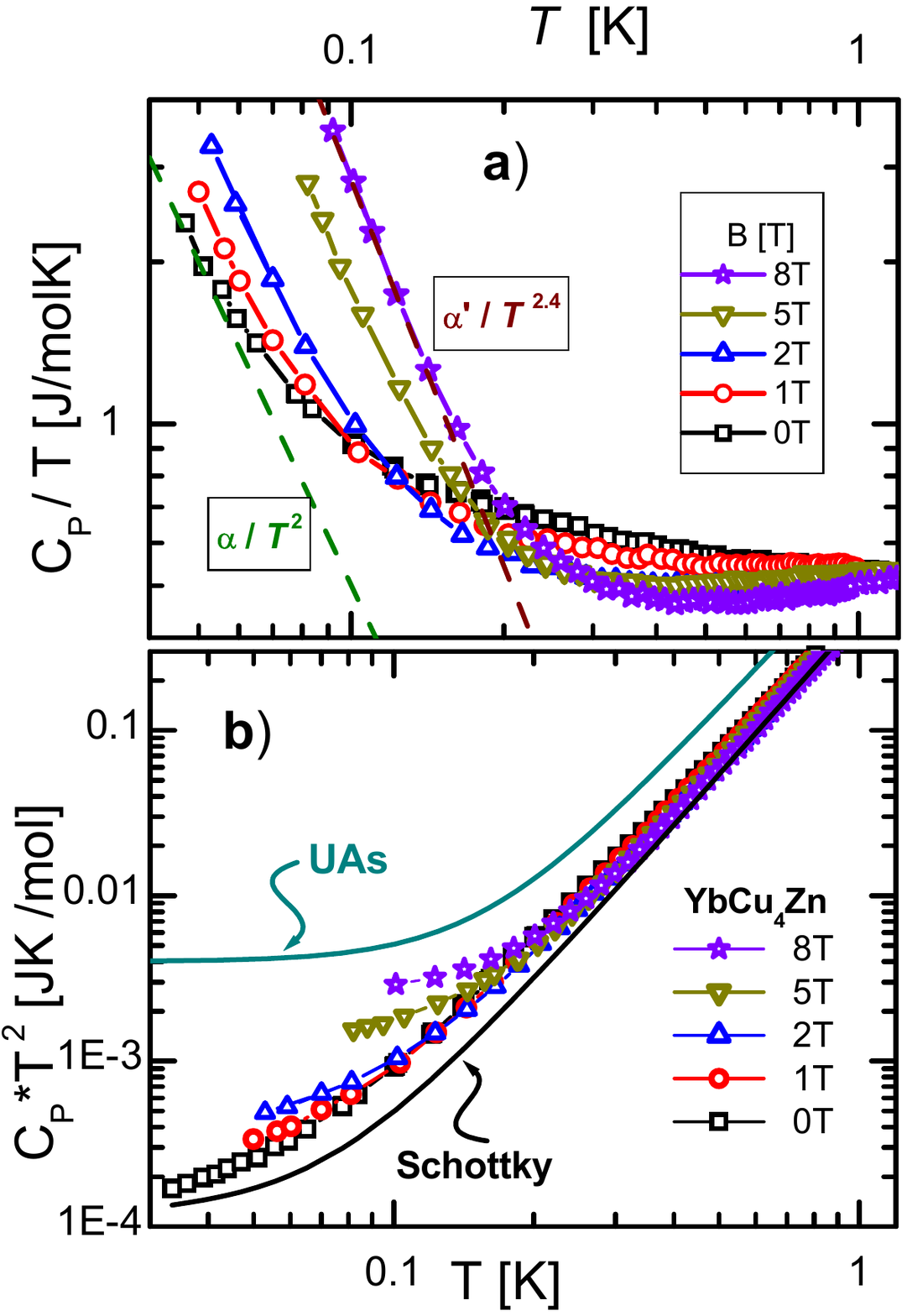}
\caption{(Color online) a) Low temperature specific heat of YbCu$_4$Zn divided temperature remarking the nuclear $C_N(T)$ contribution. Green and brown lines indicate the 
different exponent of thermal dependence: $C_N/T^n$, for $B=0$ and $B=8$\,T respectively. b) $C_P*T^2$ representation evidencing the exponent variation with 
field (see text). Continuous (black) curve is the Schottky anomaly better approaching the behavior for $B=0$. The dark-cyan curve represents the Schottky like behavior of UAs \cite{Ott} 
as a reference of a system with $C_N/T^2$. Note that in the figure the total measured specific heat $C_P/T$ is represented, that includes phonon and electronic band contributions in double 
logarithmic scale. \label{F6}}
\end{center}
\end{figure}

Paying attention, once again, to the $T\leq 0.1$\,K side of Fig.~\ref{F1}, we have remarked that |the $C_P/T$ measurements on YbCu$_4$Zn and YbCu$_{4.6}$Au$_{0.4}$ are compared with a 
$C_N/T \propto 1/T^2$ dependence and not with the expected $C_N(T))/T \propto 1/T^3$, which corresponds to the high temperature limit of a Schottky type anomaly describing the split of 
nuclear levels produced by the effective field $B_{eff}$ arising from the 
orbiting $4f$ electrons \cite{Tari}. 
In fact, as it is shown in Fig.~\ref{F5} for Au$\underline{Yb}$, the nuclear contribution is properly described by a Schottky anomaly with a $C_N \propto \alpha_n /T^2$ high temperature 
tail.
Due to the limited range in which this temperature dependence is measured, the exponent $n=2$ has to be taken as an approximation reference. Nevertheless, it is far from the 
expected value $n=3$.  
In order to reduce this uncertainty, we have also studied the $C_N(T)$ evolution as a function of field, presented in Fig.~\ref{F6}a as $C_P(T,B)/T$ for $0\geq 
B \geq 8$\,T. There one can appreciate that neither for the highest applied field the $\propto 1/T^3$ dependence is reached. 

The total measured signal includes nuclear, electronic band 
and phonon contributions: $C_P/T = C_N/T + C_{el}/T + C_{phon}/T=\alpha/T^3 +\gamma + \beta*T^2$, with a negligible contribution of the $\beta*T^2$ term in the $T<1$\,K range. 
Two reference (dashed) lines are included in the figure, one for the $B=0$ measurement as $\alpha_n/T^{2}$ (green) and the second as $\alpha_n^*/T^{2.4}$ for $B=8$\,T (brown). 
Note that the fit for $B=0$ is affected by a significant error (around 15\%) because of the few points involved. Nevertheless, the relevant physics concerns the clear difference  with the 
expected $\propto 1/T^3$ dependence, and the variation of that exponent increasing the  magnetic field.  

In order to make more evident the previously discussed change of the exponent we represent in Fig.~\ref{F6}b the temperature dependence of the specific heat as: $C_P*T^2 
= \alpha_n + \gamma*T^3$ to better visualize the deviation from the high temperature tail of the Schottky anomaly. The  $C_P*T^2$ extrapolation to $T\to 0$ ranges between 
$2\times 10^{-4}$ for $B=0$ and $3\times 10^{-3}$\,JK/mol for $B=8$\,T. These values are compared with those of UAs \cite{Ott} (black curve) included into the figure as reference and an 
'ad hoc' Schottky anomaly. Both references were chosen because they show respective $C_N \propto 1/T^2$ dependencies in a similar range of temperature like YbCu$_4$Zn and, 
consequently, a temperature independent $C_N*T^2$ contribution at $T \to 0$. 

In addition to the increase of the nuclear coefficient $\alpha_n(B)$ with field, in the figure one can appreciate how the $C_P*T^2$ data evolves from a clearly different curvature compared to 
the Schottky anomaly at $B=0$ towards that of the reference UAs shape for the $B=8$\,T.
One might claim from Fig.~\ref{F6}b that at $B=0$ the $\lim_{T \to 0}$ of $C_N*T^2$ tends to a constant $\alpha_0$, however the high temperature $T>>\Delta$ dependence of a pure 
Schottky anomaly ($\Delta$ = level splitting) starts 
to deviate from $C_N \propto 1/T^2$ around $T=\Delta$ with the onset of a negative curvature. That range of temperature is reached at higher field measurements,  
without showing such curvature. 

\begin{figure}
\begin{center}
\includegraphics[width=20pc]{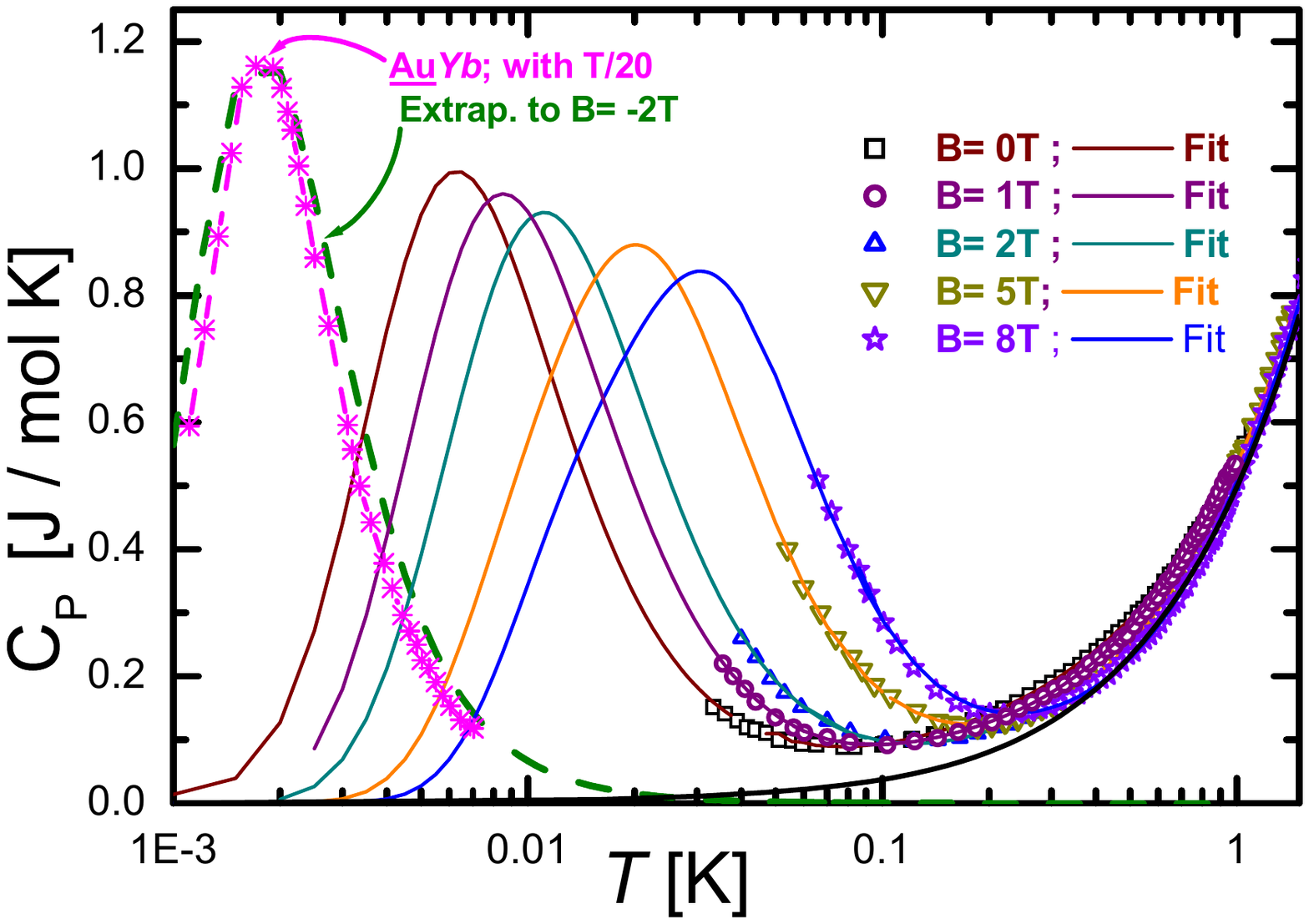}
\caption{(Color online) Fits of the specific heat using Schottky anomalies in a semilogarithmic representation, that accounts for the excitation from the singlet GS $\vec{\bf F_0}$ to the triplet 
$\vec {\bf F_1}$. A level broadening for the upper level (see text) and the electronic contribution: $\gamma *T$ are included. Green dashed 
curve is the reference for the $\gamma *T$ contribution. Magenta dashed curve represents the $C_P(T)$ of  Au${\underline{Yb}}$ after \cite{Mignot} with $T_{max}/10$, which is included as 
reference. \label{F7}}
\end{center}
\end{figure}

To gain insight into the origin of this unexpected variation of the exponent $n(B)$ in $C_N \propto 1/T^n$, we have fitted the $C_P(T)$ data with a Schottky anomaly that represents the 
excitation from the GS $\vec{\bf F_0}$ (singlet) to the excited state $\vec{\bf F_1}$ (triplet). 
To properly describe the experimental results, we have introduced a correction to the standard '$hf$' interaction, which is based on the hypothesis that a broadening of the electronic levels is 
produced by a weak remnant electronic Kondo effect, which is reflected in the hyperfine effective field as a small dispersion in the energy levels. To that purpose, the upper nuclear level 
broadening of the $ \vec {\bf F_1}$ (triplet) was modeled using a Lorentian function instead of  Dirac one.

For this analysis of the nuclear contribution we have considered that one Formula Units contains 
0.14 $^{171}$Yb isotope. 
These fits to $C_P(T, B)$ are shown in Fig.~\ref{F7} form $B=0$ up to 8\,T, including respective  electronic contributions: $\gamma *T$, which is described by the  
green dashed curve. Also for comparison, the $C_P(T)$ of  Au${\underline{Yb}}$ after \cite{Mignot} is included, with a scaling of its by $T_{max}/10$, see magenta dashed curve.

\begin{figure}
\begin{center}
\includegraphics[width=20pc]{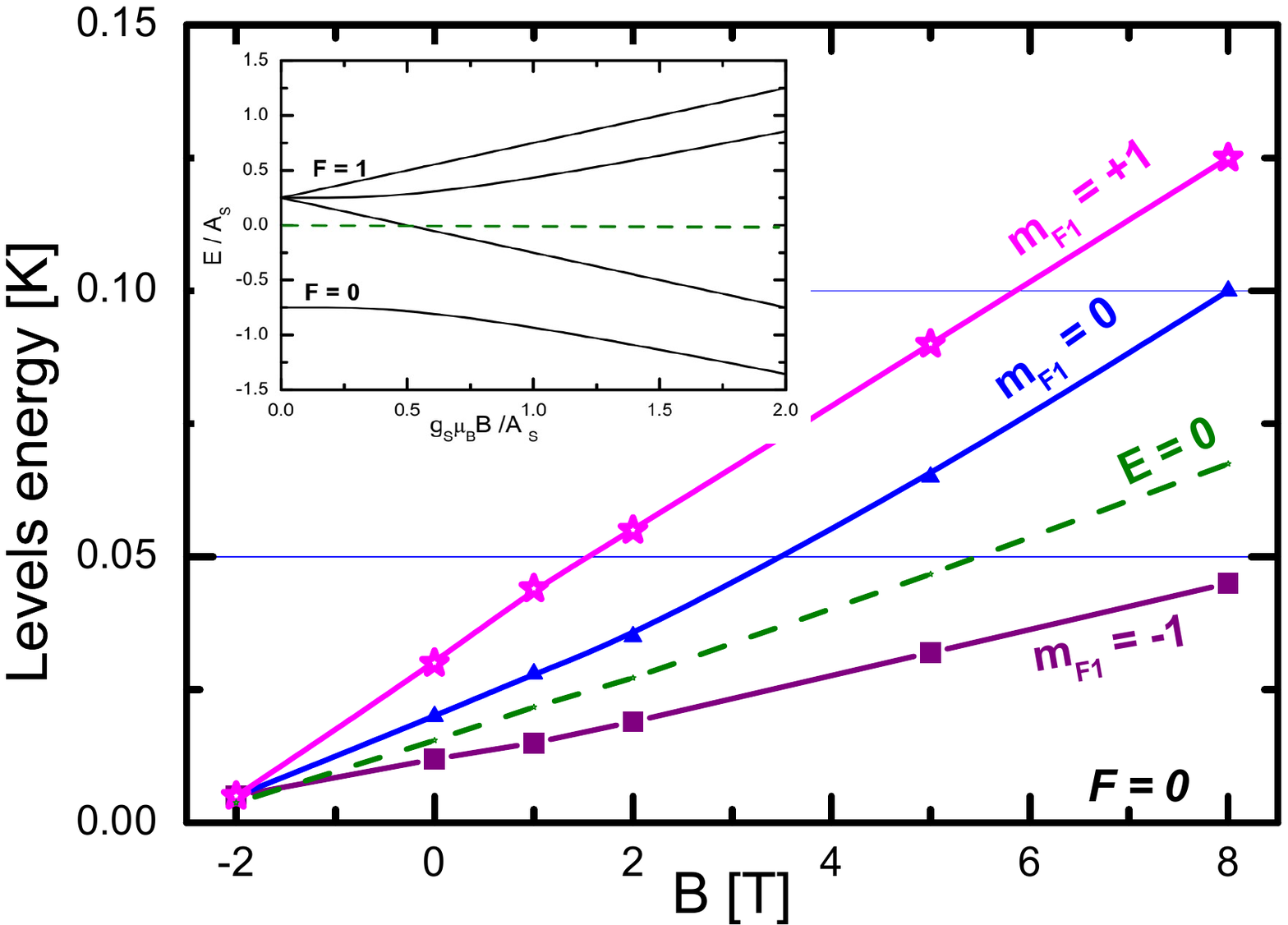}
\caption{(Color online) Energy variation of the $\vec {\bf F_1}$ (triplet) levels by magnetic field, measured from the  $\vec{\bf F_0}$ (singlet) GS. Green dashed line:  
baricenter of energy of the full system ($E_0$). Inset: theoretic evaluation of the single-triplet system as a function of field, taking as reference the energy baricenter $E_0$. \label{F8}}
\end{center}
\end{figure}

The level energies of the excited triplet with projections: $m_{\vec {\bf F_1}}=+1$, $m_{\vec {\bf F_1}}=0$ and $m_{\vec {\bf F_1}}=-1$, are presented in Fig.~\ref{F8} as a 
function of field. The zero energy reference corresponds to the singlet-GS $\vec{\bf F_0}$.
The green dashed line 
indicates the energy variation of the baricenter ($E_0$) of the full set of nuclear levels . Note that the  $m_{F1}$ levels are already split at $B=0$ by an internal $H_{hf}$ field. The convergence of 
those levels, i.e. $H_{hf}$=0 occurs at $B=-2$\,T. 
The theoretic evaluation of the single-triplet sytem as a function of field is included for comparison in the inset, where the baricenter of the energy: $E=0$, is taken as energy 
reference. 

Since the range of temperature of the measured specific heat only covers the upper tail of the Schottky anomaly, for practical reasons only the broadening ($\delta m$) of the highest energy 
level:  $m_{F1}=+1$, of the $\vec {\bf F_1}
$ (triplet) was computed to fit the $C_P(T, B)$ curves shown in  Fig.~\ref{F7}. 
Within the statistical error of these fits we consider that this upper level contains the most relevant information related to the physics under study.   
The computed value is: $\delta m_{+1}=25 \pm 5$\,mK, which  
slightly decreases with increasing field. 

The energy related to the broadening of the nuclear level:  $\delta m_{+1}=25 \pm 2$\,mK, can be compared with the Kondo temperature: $T_K=6.1$\,K, of the electronic GS in terms of the 
effective fields: $H_{T_K}=k_B T_K/g_{eff}\mu_B$ and $H_{\delta_m}=k_B \delta_m/g_N\mu_N$   respectively. Where $g_{eff}=0.6$ as it was evaluated from $M vs B$ at 2\,K in 
the lower inset of Fig.~\ref{F4}, and $g_N=3.34$ taken from {\underline {Au}}{\it Yb}  \cite{gNTao}, being $g_{eff}/g_N=0.6/3.34=0.17$ and $\mu_B/\mu_N=1836$. Thus, the ratio: 
$H_{T_K}/H_{\delta_m}= (k_BT_K/g_{eff}\mu_B)/(k_B\delta_m/g_N\mu_N)$ = (6.1/0.025)/(0.17*1836) =0.78, indicates coincident values within the approximations introduced in this 
evaluation. 

Therefore, the deviation from the expected 
$C_N\propto 1/T^2$ dependence can be attributed to the effect of the Kondo interaction on the effective electronic field perceived by the nuclear levels.

\section{Conclusions}

We have confirmed that YbCu$_4$Zn is tuned very close to a QCP after its NFL behavior of  $C_m/T\propto -log(T/T_Q)$ dependence and the low values of the characteristic energies 
extracted from thermal and magnetic properties like: quantum fluctuations $T_Q = 6$\,K and  Kondo temperture $T_K^{GS} =\theta_P/\sqrt 2= 6.1$\,K.

From the comparison of respective physical properties of the YbCu$_4$M (M= Ni, Au and Zn) family and the isotipic YbNi$_4$Mg, one observes that some properties, e.g. $\chi(T\to 0)$, are 
similar on both sides of the QCP. However, a clear difference appears in the $C_P(T)$ between power laws and logarithmic ones. 
This reveals that, though the energy scales are similar, the excitation spectrum is clearly different. This set of six compounds does not prove a general behavior, nevertheless to our 
knowledge it is the largest group of isotipic compounds tuned in the vicinity of a QCP.

The relevant coupled hyperfine states are $\vec {\bf F} = \vec {\bf I} \pm \vec {\bf S}  = 0$  and 1 according to nuclear $\vec {\bf I}$ = 1/2 and electronic $\vec {\bf J}$ = 1/2, with the  $
\vec {\bf F_0}$ singlet as a GS and $\vec {\bf F_1}$ triplet as excited state. Nevertheless, the $\vec {\bf F_1}$ triplet is already split at $B=0$ by an hyperfine field $H_{hf} \approx 
2$\,T. The field effect induces a Zeeman-like increase of $A_{hf}$ from $\approx 25$\,mK at  $B=0$ to $\approx 130$\,mK for $B=8$\,T

Notably, the nuclear contribution of YbCu$_4$Zn: $C_N(T,B) = 1/T^n$ does not obey the expected $n=2$ dependence but a fraction with $1 \leq n \leq 1.4$ for $0 < B < 8$\,T. Therefore the fits of $C_N(T,B)$ require to account for a nuclear levels broadening of $\approx 5$\,mK. 
This indicates that the Kondo mechanisms acting on the $4f$ electrons is reflected in the  hyperfine coupling producing a concomitant energy widening of the nuclear levels.

The weakness of this analysis is the short range of temperature within which the $C_N(T,B)$ fit can be performed, however the physical description is strengthened by using the same fitting 
formula for 5 different magnetic fields. Independently of the significant indeterminacy in the value of the exponent $n$, the relevant information corresponds to its variation with field, 
that tends to the expected value $n=2$ once the electronic Kondo effect is quenched by the applied magnetic field.

\end{document}